# Graphene-Supported Silver-Iron Carbon Nitride Derived from Thermal Decomposition of Silver Hexacyanoferrate as Effective Electrocatalyst for the Oxygen Reduction Reaction in Alkaline Media

Beata Dembinska[1*], Kamila Brzozowska[1], Adam Szwed[1], Krzysztof Miecznikowski[1], Enrico Negro[2], Vito Di Noto[2], Pawel J. Kulesza[1*]

[1]*Faculty of Chemistry, University of Warsaw, Pasteura 1, PL-02-093 Warsaw, Poland*
[2]*Department of Industrial Engineering, Università degli Studi di Padova in Department of Chemical Sciences, Via Marzolo 1, 35131 Padova (PD) Italy*

___________

*Corresponding Authors

E-mail: *bbaranowska@chem.uw.edu.pl*; Tel.: +48225526336

E-mail: *pkulesza@chem.uw.edu.pl*; Tel: +48225526344

**Abstract**

Silver-iron carbon nitride has been obtained by pyrolysis (under inert atmosphere) of silver hexacyanoferrate(II), precipitated on graphene nanoplatelets, and examined as electrocatalyst for oxygen reduction reaction in alkaline media in comparison to silver nanoparticles and iron carbon nitride (prepared separately in a similar manner on graphene nanoplatelets). The catalytic materials have been studied in 0.1 M potassium hydroxide electrolyte using such electrochemical diagnostic techniques as cyclic voltammetry and rotating ring-disk electrode voltammetry. Upon application of graphene nanoplatelets-supported mixed silver-iron carbon nitride catalyst, the reduction of oxygen proceeds at more positive potentials and the amounts of hydrogen peroxide (generated during reduction of oxygen at potentials more positive than 0.3 V) are lower relative to silver nanoparticles and iron carbon nitride (supported on graphene nanoplatelets) examined separately. Promoting effect is ascribed to high activity of silver toward the reduction/decomposition of $H_2O_2$ in basic medium. Additionally, it has been observed that the systems based on carbon nitrides show considerable stability due to strong fixation of metal complexes to CN shells.

**Introduction**

For several past decades Proton Exchange Membrane Fuel Cells (PEMFCs), as alternative environmentally-friendly devices for energy conversion, have dominated the area of development of low temperature fuel cells due to advanced technology of manufacture of cation conductive membranes. But in the last few years substantial progress in the preparation of anion exchange membranes have driven enormous expansion of research work on Alkaline Fuel Cells (AFCs) [1,2]. Faster kinetics of electrocatalytic processes in alkaline media in relation to acidic electrolytes [3,4] makes it possible to substitute platinum and platinum-group metals with more abundant and less expensive materials like silver [4-9], transition metals (especially Fe, Co) in coordination with nitrogen [10-12], transition metal oxides [13,14] and even carbon nanostructures doped with heteroatoms (N, S, P, B) [15-18].

The most widely examined group of oxygen reduction catalysts in alkaline media include carbon-supported iron coordinated with nitrogen (abbreviated as *FeNC*) obtained through high temperature pyrolysis of simple precursors. The active sites are believed to resemble the structure of the centers of macromolecular complexes present in porphyrins or phtalocyanines. An interesting approach developed recently comprises carbon nitride-based electrocatalysts prepared from thermal decomposition of hexacyanometalates where carbon-based matrix embeds nitrogen atoms and coordinates metallic species [19-27]. The advantage of using hexacyanometalates as starting compounds is the possibility of synthesis of materials with controlled stoichiometry and also containing mixed metallic species (e.g. Fe, Co, Ni, Sn, Mn, Cu, Ag).

Silver has been considered as peculiar promising alternative metal (to platinum) due to its low cost and considerable electrocatalytic activity and stability toward both oxygen and hydrogen peroxide reduction reactions [4-9,28,29]. The mechanism of oxygen reduction on Ag usually is reported to proceed nearly directly to water. What is more, structural dependence of the kinetics of oxygen reduction proceeding at Ag was also revealed; it increases in the order (100) < (111) < (110) planes [30]. In addition, recently highly active and stable materials have been obtained through the combination of silver with manganese oxides [31,32], cobalt oxides [33], molybdenum oxides [34] and other transition metals (Co, Cu, Ni, Sn, Fe) [35-39].

Cathode operating in alkaline fuel cells requires carbon support which, except high electrochemically accessible surface area and suitable porosity for proper mass transport, is characterized by high graphitization degree which determines the electron conductivity and stability of the whole electrocatalytic system. In this respect, graphene-related materials with their unique properties could theoretically enable not only the increase of the effectiveness of catalysts but also their stability when compared to materials based on conventional carbon blacks [40,41]. However, the lack of surface functional groups and defects makes the pristine graphene difficult material for further operation including the synthesis of homogenously distributed catalytic materials. Therefore, in most studies reduced graphene oxide (rGO) or graphene oxide (GO) have been used as supports or their precursors which indeed allow obtaining finely dispersed nanoparticles, but the stability of such materials and electronic conductivity is usually reported to be insufficient [41,42]. On the other hand, it has recently been proven that high durability can be obtained using polyelectrolyte-modified graphene nanoplatelets (GNP) as support for Pt nanoparticles which was attributed to the intrinsic high graphitization degree of GNP and the enhanced Pt-carbon interaction in Pt/GNP [43]. Nevertheless, irrespective of graphene-related materials used, the problem of gas permeability and water management (due to stacking effects) leads to lower intrinsic performances of cathodes during oxygen reduction reaction in relation to those based on other carbon nanostructures. In fact graphene and its derivatives are commonly used as effective additives to gas-barrier systems [44]. The problem can be overcome by fabrication of composite assemblies with other carbonaceous materials acting as spacers between the graphene-type sheets, thus creating three-dimensional morphologies [41,42,45-47]. In addition, it is well established that the durability of such hybrid systems can be significantly improved [41,42,45-47].

As a result of extensive work concerning graphene-related materials (including nitrogenated derivatives) there have also been promising attempts to utilize them as supports or modifiers for silver based catalysts active toward oxygen reduction in alkaline environment [48-51]. Therefore, in the present work we report the synthesis and characterization of graphene nanoplatelets-supported silver-iron carbon nitride derived from thermal decomposition of silver hexacyanoferrate(II) deposited onto polycarboxylate functionalized GNP. The material is employed as electrocatalyst for oxygen reduction reaction in 0.1 M KOH and its activity is compared to silver nanoparticles and iron carbon nitride (prepared separately in a similar manner on GNP). The stability issue of examined systems, due to the presence of strong fixation of metal complexes to CN shells, is also discussed.

**Experimental**

All chemicals were commercial materials of the highest available purity. $HClO_4$, ethanol, 2-propanol, KOH, $K_4[Fe(CN)_6] \cdot 3H_2O$ were from POCh (Poland), the solution of 5% Nafion-1100, $AgNO_3$, $Fe_2(SO_4)_3 \cdot 6H_2O$ and polycarboxylate functionalized graphene nanoplatelets (GNP) from Sigma-Aldrich, Vulcan XC-72R (C) from Cabot (USA), graphene oxide sheets (GO) of 300–700 nm sizes (thickness, 1.1 ± 0.2 nm) from Megantech, Pt(20%)/C from E-Tek, hydrogen peroxide 30% from Chempur (Poland) and nitrogen, oxygen and argon gases (purity 99.999%) are from Air products (Poland). The solutions are prepared from doubly-distilled and subsequently-deionized (Millipore Milli-Q) water.

***Ag$_4$HCF/GNP*** sample was prepared by precipitation method. In the procedure the mixture of 15.1 ml of water, 2 ml of ethanol and 50 mg of GNP was placed for 1 h in an ultrasonic bath to obtain its good dispersion. Then 10.3 ml of 10 mM AgNO$_3$ water solution was added and the slurry was left in the bath for about 15 minutes followed by mixing at magnetic stirrer. After 10 min of mixing 2.6 ml of 10 mM K$_4$[Fe(CN)$_6$] water solution was added and the dispersion was left for several hours to assure full precipitation of the product (white sediment in a blank test). Total volume of mixture was 30 ml. Next the solution was centrifuged and removed. The sediment was washed three times with water and dried on a hot plate at about 60 °C. Considering the amounts of GNP and precursors used for synthesis and assuming that the main product of the reaction is Ag$_4$HCF overall content of Ag and Fe in the sample was about 20 % by weight relative to GNP.

***Fe$_4$HCF$_3$/GNP*** sample was synthesized in an analogous manner as Ag$_4$HCF/GNP, but, in order to obtain insoluble form of Prussian Blue, $Fe^{3+}$ was used in excess. Therefore during the synthesis 7.4 ml of 10 mM solution of Fe$_2$(SO$_4$)$_3$ and 9.6 ml of 10 mM K$_4$[Fe(CN)$_6$] were added. The amount of water and ethanol solutions were adjusted to maintain 30 ml of the total volume of the mixture. Product (dark blue sediment in a blank test) was washed and dried. The content of Fe in the sample was about 20 % by weight relative to GNP.

***AgNO$_3$/GNP*** sample was prepared by simple impregnation method: to 18.4 ml of homogenized dispersion of GNP in water-ethanolic solution 11.6 ml of 10 mM AgNO$_3$ was added. The slurry was left at magnetic stirrer for several hours, and eventually solvent was evaporated. The content of Ag in the sample was about 20 % by weight relative to GNP.

All prepared samples were grinded in an agate mortar and annealed in a silica tube under Ar atmosphere at 800 °C for 2 h (heating up rate of the oven was 4 °C min$^{-1}$). Heat-treated samples are labeled **AgFeCN$_x$/GNP800**, **FeCN$_x$/GNP800** and **Ag/GNP800**. For comparative studies GNP was also washed with water and annealed under the same conditions as described above (it is labeled **GNP800**) and additional sample of **AgFeCN$_x$/GO800** was prepared.

The electrochemical measurements were performed with CH Instruments (Austin, TX, USA) Model 750D and 920D workstations. All electrochemical experiments were conducted at room temperature, 22 ± 2 °C. Mercury-mercury sulfate electrode (Hg/Hg$_2$SO$_4$) (in acid media) and saturated calomel electrode (SCE) (in alkaline media) were used as reference electrodes and all potentials are expressed against the RHE. Glassy carbon rod or platinum sheet served as counter electrodes. Rotating ring-disk electrode (RRDE) experiments were performed using a variable speed rotator (Pine Instruments, USA). The electrode assembly utilized a glassy carbon disk (with geometric area of 0.247 cm$^2$) and Pt ring. The collection efficiency (N) of the RRDE assembly was equal to 0.388, as determined from the ratio of ring and disk currents (at 1600 rpm) using the argon-saturated 0.005 mol dm$^{-3}$ K$_3$[Fe(CN)$_6$] in 0.1 mol dm$^{-3}$ K$_2$SO$_4$ solution. Before electrode layers preparation working electrodes were polished with aqueous alumina slurries (grain size, 5-0.05 μm) on a Buehler polishing cloth. In the course of RRDE experiments, in order to oxidize H$_2$O$_2$ (generated at the disk) under convection-diffusional control, the potential of the ring electrode was kept at 1.23 V vs. RHE.

Electrode layers were deposited on a glassy carbon disk by dropping appropriate volumes of homogenized inks containing 5 mg of nanomaterial (catalyst grinded in an agate mortar with Vulcan® XC-72R (labeled as **C**) in the mass ratio of 1 to 1), 500 μl of solvent (2–propanol) and 6 μl of binder (5% solution of Nafion®) and subsequently dried in room temperature, 22±2 °C. Final content of Nafion® was 5% in relation to the weight of catalyst and Vulcan. Catalysts loadings were 600 μg cm$^{-2}$ and in the case of the Pt(20%)/C, the Pt loading was 15 μg cm$^{-2}$. The electrodes covered with the catalytic layers were copiously washed out with the stream of water (in order to clean the surface from impurities) and subjected to cycling at a sweep rate of 100 mV·s$^{-1}$ in nitrogen-saturated 0.1 mol dm$^{-3}$ KOH in the potential range from 0.05 V to 1.05 V to activate the layers and receive reproducible responses.

In the experiments involving preconditioning step (labeled *P*), the electrode was cycled in nitrogen-saturated 0.1 M HClO$_4$ solution in the range of 0.05 - 1.05 V at a sweep rate of 100 mV·s$^{-1}$ until stable (flat) response was obtained. The electrolyte was changed two times during the experiments.

In the additional experiments conducted in 0.5 M K$_2$SO$_4$, concerning electrochemical characteristics of precursors i.e. Fe$_4$HCF$_3$/GNP, Ag$_4$HCF/GNP and GNP carrier, inks were prepared without Vulcan additive and the loadings of electrode layers were 300 μg cm$^{-2}$.

Transmission Electron Microscopy (TEM) experiments were carried out with Libra 120 EFTEM (Carl Zeiss) operating at 120 kV. Scanning electron microscopic (SEM) measurements and energy-dispersive X-ray analysis were achieved using MERLIN FE-SEM (Carl Zeiss) equipped with EDX analyzer (Bruker). X-Ray diffraction (XRD) spectra were collected using Bruker D8 Discover equipped with a Cu lamp (1.54 Å) and Vantec (linear) detector.

**Results and discussion**

Fig. 1 illustrates representative SEM and TEM images of bare GNP800 support and after its modification with catalytic nanoparticles. It is evident from Fig. 1A and B that the GNP support indeed exists in the form of platelets with the sizes ranging from few hundred nanometers up to few micrometers. In a case of Ag/GNP800 (Fig. 1. C and D), silver forms large particles with diameters from 50 nm to 200-300 nm; virtually no smaller species could be observed in the series of images. TEM images of precursor (AgNO$_3$) recrystallized onto GNP (data not shown) revealed that polycarboxylate-modified GNP enabled good dispersion of crystallites, with diameters from few to tens of nm (but usually they did not exceed 100 nm). It is reasonable to expect that the aggregation occurs under high temperature (800 ºC) of annealing. In contrast, the micrographs of FeCN$_x$/GNP800 (Fig. 1E and F) and AgFeCN$_x$/GNP800 (Fig. 1G and H) show that except particles with diameters of hundreds of nm there is also appreciable population of much smaller species. Therefore, it may be concluded that the application of the precursors in the form of such inorganic metal clusters as Prussian blue and its analogs lowers (at least to a certain degree) sintering of metal nanoparticles. What is more, it is also evident that in the FeCN$_x$/GNP800 (Fig. 2F) and AgFeCN$_x$/GNP800 (Fig. 2H) samples, metal nanoparticles are embedded in a shell, which thickness the most probably depends on stoichiometry of starting materials. The results stay in accordance with previous reports [22-26] where similarly obtained nanomaterials consisted of metal nanoparticles encapsulated in a compact carbon nitride shell (CN).

The EDX analysis of the materials is generally consistent with the amounts of precursors taken to the synthesis of materials. In the case of Fe$_4$HCF$_3$/GNP the molar ratio of Fe to N is 0.5 (which is close to the value 0.4 expected from stoichiometry of the compound) and after pyrolysis (sample FeCN$_x$/GNP800) it is substantially changed exceeding the value of 0.1 implying substantial loss of nitrogen species. The data obtained for Ag$_4$HCF/GNP (before annealing) reveal that the molar ratio of the elements Ag : Fe : N is 3 : 1 : 6 and for AgFeCN$_x$/GNP800 (after annealing) the composition change into the following: 3 : 1 : 3, again implying the loss of N. The analytical data also revealed the presence of carbon, oxygen as well as some content of potassium. The EDX mapping of FeCN$_x$/GNP800 (Fig. 2A) shows rather homogenous distribution of N and Fe. In a case of AgFeCN$_x$/GNP800 (Fig. 2B) it is evident that the signals originating form Fe and Ag are concentrated in some sites and it seems that large portion of them superimpose which may reflect the mixed nature of obtained nanoparticles.

As shown in Fig. 3a, XRD pattern of Ag/GNP800 presents peaks characteristic of face centered cubic (fcc) crystal structure of Ag ((111), (200) and (220)) with cell parameter 4.083 Å and average crystallite size of 51.7 nm (calculated from the Scherrer equation). In the XRD spectra of $FeCN_x$/GNP800 (Fig. 3b) the main reflections correspond to body centered cubic structure (bcc) of α-Fe ((110) and (200)) with cell parameter 2.862 Å and average crystallite size of 32.7 nm. Some other, but much less distinct, peaks in Fig. 3b could be ascribed to the traces of oxide forms of iron, like γ-$Fe_2O_3$ or $Fe_3O_4$. One has to be aware that for Fe containing materials the fluorescence of X-rays, arising when Cu lamp is used, increase the background noise. In the case of $AgFeCN_x$/GNP800 (Fig. 3c), except signals characteristic of Ag and α-Fe, there are also small peaks probably coming from the traces of iron oxides. Here the cell parameter for Ag is 4.075 Å (average crystallite size of Ag is 92.8 nm) and for Fe - 2.886 Å (average crystallite size of Fe is 104.6 nm). It is worth to emphasize that the Sherrer's formula, as an approximation, gives reliable results for nearly spherical particles. In all patterns GNP support reflections are present ((002), (111) and (004)) indicating partial stacking.

The presence of the corresponding pristine compounds (silver and iron(III) hexacyanoferrates(II)) on the surface of GNP was examined by cyclic voltammetry experiments conducted in 0.5 M $K_2SO_4$. Fig. 4A shows voltammograms recorded for GC electrode covered with $Fe_4HCF_3$/GNP (a) and $Ag_4HCF$/GNP (b) compared to GNP support (c). Polarization curve recorded for $Fe_4HCF_3$/GNP reveal well known behavior of PB: its' oxidation to iron(III) hexacyanoferrate(III), at about 1.2 V, and reduction to iron(II) hexacyanoferrate(II), at about 0.4 V [52]. In a case of $Ag_4HCF$/GNP, after first few scans, where oxidation of hexacyanoferrate(II) into hexacyanoferrate(III) coupled with the loss of $Ag^+$ into the bulk solution is observed, voltammogram resembles typical characteristics of silver hexacyanoferrate(III) [53].

Fig. 4B presents cyclic voltammetric responses of (a) Ag/GNP800-C, (b) FeCN$_x$/GNP800-C, (c) AgFeCN$_x$/GNP800-C and (d) GNP800-C layers deposited on glassy carbon electrode recorded in the de-aerated 0.1 M KOH. Except FeCN$_x$/GNP800-C (b), in applied range of potentials, all catalysts are characterized by similar capacitive currents. For FeCN$_x$/GNP800-C a gradual rise in redox responses, observed at lower potentials, emerge due to the presence of higher amount iron species when compared to other compounds.

Fig. 5A illustrates background subtracted (normalized) RDE responses (backward scans) recorded in the oxygen-saturated 0.1 M KOH (at 1600 rpm) at the disk electrode covered with catalytic films. Data indicate that all three catalysts: Ag/GNP800-C (Curve a), FeCN$_x$/GNP800-C (Curve b) and AgFeCN$_x$/GNP800-C (Curve c) are much more active than bare GNP800-C carrier (Curve d) in terms of higher catalytic currents and lower overpotentials of the oxygen reduction process. What is more, the onset and half-wave potentials of oxygen reduction are slightly shifted toward more positive values for AgFeCN$_x$/GNP800-C relative to both Ag/GNP800-C and FeCN$_x$/GNP800-C, indicating enhanced activity of the sample. It is also evident that, while for Ag containing samples (Curves a and c) the convective-diffusion-limiting current plateaus are reached (at about 0.6 V), in the case of FeCN$_x$/GNP800-C (Curve b) plateau is followed by further increase of the currents (from about 0.5 V). At this point it should be emphasized that complications in the interpretation of oxygen reduction RDE curves at lower potentials in alkaline media may arise due to the competition for $O_2$ between metallic centers and different forms of carbon (existing here as GNP, Vulcan additive and CN shell) which are also active towards oxygen reduction [29]. Eventually, it can be also observed that the half-wave potential for oxygen reduction at the AgFeCN$_x$/GNP800-C catalyst (~0.71 V) is still ca. 120 mV less positive than that characteristic of conventional carbon-supported platinum (~0.83 V) (Fig. 5A, Curve e). Yet, one has to be aware of the fact that comparison to conventional Pt/C (containing 3-4 nm Pt nanoparticles) is not straightforward here because our catalysts, due to synthetic approach, are much larger. The preparation of our samples included long-term high temperature pyrolysis, necessary to generate CN shells, and under such conditions metallic nanoparticles undergo inevitable aggregation. In the future, to overcome the drawback, rapid heating and shortening the period of pyrolysis could be more suitable route for fabrication of smaller particles for fuel cell applications.

It is apparent from Fig. 5B (where the amount of $HO_2^-$ formed during electroreduction of oxygen is monitored at the platinum ring electrode) that Ag/GNP800-C (Curve a), $FeCN_x$/GNP800-C (Curve b) and $AgFeCN_x$/GNP800-C (Curve c), although less selective toward four electron reduction of oxygen than standard Pt(20%)/C (Curve e), exhibit much lower currents corresponding to oxidation of the $HO_2^-$ intermediate in relation to unmodified GNP800-C carrier (Curve d). Data also indicate that for Ag containing samples (Curves a and c in Fig. 5B) $HO_2^-$ oxidation currents, in the potential region above 0.3 V, are considerably decreased in comparison to only iron metal containing catalyst (Curve b in Fig. 5B).

Upon consideration of both $HO_2^-$ oxidation (ring) and $O_2$ reduction disk currents (Fig. 5A and B), the relative (percentage) formation of intermediate product and the corresponding number of transferred electrons ($n_e$) per oxygen molecule involved in the oxygen reduction were calculated using the following equations [5,54,55]:

$$X_{HO_2^-} = \frac{200 I_R/N}{I_D + I_R/N} \qquad \text{Eq. (1)}$$

$$n_e = \frac{4 I_D}{I_D + I_R/N} \qquad \text{Eq. (2)}$$

where $I_R$ is the ring current, $I_D$ is the disk current and N is the collection efficiency of the RRDE assembly. Fig. 5C shows $X_{HO_2^-}$ plotted versus potential applied to the disk electrode. It can be observed from Fig. 5C that, at potentials higher than 0.3 V (i.e. in the region of cathode operation in a fuel cell), the fraction of produced $H_2O_2$ is lower (up to 20%) for AgFeCN$_x$/GNP800-C (Curve c) relative to both Ag/GNP800-C (up to 25%, Curve a) and FeCN$_x$/GNP800-C (up to 47 %, Curve b). In addition, as expected, at unmodified GNP800-C carrier (Fig. 5C, Curve d) relative amount of $H_2O_2$ intermediate is on the level of 90% and at standard Pt(20%)/C (Fig. 5C, Curve e) – 4%. The dependence of number of exchanged electrons ($n_e$) plotted versus disk potential (Fig. 5D) reveals that at potentials close to 0.5-0.6 V nearly three-electron reduction mechanism is the dominating pathway for the oxygen reduction at FeCN$_x$/GNP800-C. Definitely higher values of $n_e$ are observed for silver-containing systems, wherein the parameter for AgFeCN$_x$/GNP800-C the most closely approaches the one observed for standard Pt(20%)/C. The results presented in Fig. 5 imply that silver is effective component of catalytic materials in terms of increasing their activity and selectivity towards 4-electron reduction of oxygen in alkaline environment, which, together with higher abundance of the element and lower price than platinum, makes it attractive alternative for the applications in alkaline fuel cells.

In further experiments the electrochemical characteristics and electrocatalytic activity of catalysts was tested after treatment in acid media (preconditioning step, P). The following materials: Ag/GNP800-C, FeCN$_x$/GNP800-C and AgFeCN$_x$/GNP800-C were subjected to continuous potential cycling in nitrogen-saturated 0.1 M HClO$_4$. For all samples there could be observed gradual decline of currents characteristic of silver and iron species (for convenience data not shown) implying their dissolution into the bulk of the electrolyte. The experiments were terminated when no further changes in electrochemical responses were observed. Fig. 6 present cyclic voltammograms of GC electrodes covered with respective films recorded in 0.1 M KOH solution (after washing them with water). It is evident that signals coming from silver oxides/hydroxides formation and reduction (Figs 6A and C) as well as iron redox transitions (Fig. 6B) are much suppressed after preconditioning in perchloric acid (solid lines) when compared to untreated samples (dashed lines) which confirms significant loss of metallic species in all catalysts. Important observation is that the extent of changes varies from sample to sample which is particularly apparent in the case of Ag/GNP800-C (Fig. 6A, solid line) where flat background is observed and AgFeCN$_x$/GNP800-C (Fig. 6C, solid line) where small responses of silver redox peaks (additionally split) are still present after preconditioning. Two more important pieces of information can be drawn from the comparison of Figs 6A and C. The first one is the positive shift of Ag redox peaks in the sample AgFeCN$_x$/GNP800-C, especially cathodic one (of c.a. 25 mV) which may imply that more reduced state of silver is preferred relative to Ag/GNP800-C sample due to possible partial charge density transfer from Fe to Ag or changes in particle size of Ag [36]. On the other hand it has also been postulated in the literature that voltammetric characteristics of silver depends (among others) on Ag loading and surface structure [37,56]. The other information, based on the charges under oxidation/reduction signals of Ag, suggest that in AgFeCN$_x$/GNP800-C catalyst, the active electrochemical surface area of silver is about 50% of the one observed for Ag/GNP800-C. This may be explained by superposition of few factors: somehow lower amount of Ag (according to synthesis procedure described in Experimental Section), higher average crystallite size (revealed by XRD analysis) and partial covering of catalytic centers by CN shells which (as it is shown in Fig. 1H) tightly surround Ag or Ag-Fe nanoparticles in AgFeCN$_x$/GNP800-C.

Electrochemical experiments aiming at evaluation of catalytic activity and selectivity in the oxygen reduction of Ag/GNP800-C-P, FeCN$_x$/GNP800-C-P and AgFeCN$_x$/GNP800-C-P were conducted under identical conditions as in the case of untreated samples and comparative results are presented in Fig. 7. It is evident that the performance of the $O_2$ reduction for FeCN$_x$/GNP800-C-P is only slightly lowered with respect to FeCN$_x$/GNP800-C (Fig. 7B, solid and dotted lines, respectively). As far as AgFeCN$_x$/GNP800-C-P is concerned, the decrease in effectiveness toward the process is somehow more pronounced in relation to FeCN$_x$/GNP800-C (Fig. 7C, solid and dotted lines, respectively). But the greatest change (drop) in activity is induced in the case of Ag/GNP800-C-P in comparison to its untreated counterpart (Fig. 7A, solid and dotted lines, respectively). In fact its performance and selectivity toward $O_2$ reduction closely approaches that characteristic for bare GNP800-C (please compare Fig. 5A and B, Curves d). Obtained results allow to conclude that, contrary to GNP-supported silver nanoparticles, prepared by simple impregnation method, application of such inorganic coordination compounds as Prussian blue and its silver analogue as precursors enable to preserve the active sites even after preconditioning in an acid electrolyte. The increased stability should be ascribed to the existence of catalytic centers (iron and silver) in the form of coordination complexes strongly bonded to the surface of carbon nitride shell as suggested previously in refs. [22-26].

For comparative purposes we have also performed additional diagnostic experiments utilizing the sample of AgFeCN$_x$ deposited onto Graphene Oxide (GO), prepared in an analogous manner as presented above for AgFeCN$_x$/GNP800. We were expecting that GO, containing high abundance of oxygen functionalities and surface defects, would be a proper candidate for obtaining well-dispersed catalyst in spite of using high pyrolysis temperature during synthesis. Additionally, cyanides, released upon decomposition of hexacyanoferrates, are known as reducing agents, therefore their presence should at least partially reduce GO to better conducting reduced graphene oxide (rGO). Since physicochemical properties of utilized GO have already been presented in details [57], here only brief TEM characteristics of AgFeCN$_x$/GO800 and its electrocatalytic performance toward O$_2$ reduction in relation to AgFeCN$_x$/GNP800 is shown. As evident form Fig. 8A, the catalytic material AgFeCN$_x$/GO800 is characterized by fairly uniform distribution of crystals (with diameters from about 50 nm or even less up to about 100-150 nm) over support. However, the comparison of RRDE data (Fig. 8B and C) demonstrates that the activity and selectivity of AgFeCN$_x$/GO800-C, despite of the presence of Vulcan as "separator" of graphene sheets, is much weakened when compared to AgFeCN$_x$/GNP800-C. Most probably the O$_2$ and water management are still largely hindered in the former case or the electron conductivity is not sufficient enough. The results clearly imply that for electrocatalytic purposes the graphene derivatives used as supports should be selected with caution.

**Conclusions**

It is demonstrated for the first time that thermal decomposition under the inert atmosphere of silver hexacyanoferrate(II), deposited on graphene nanoplatelets, generates nanomaterial containing silver and iron species coordinated to carbon nitride shell of remarkable efficiency toward oxygen reduction process in alkaline environment. An important observation is not only the fact that at graphene nanoplatelets-supported mixed silver-iron carbon nitride the reaction proceeds at more positive potential values in comparison to both silver nanoparticles and iron carbon nitride (prepared separately under the same conditions on graphene nanoplatelets carriers) but also that the amounts of hydrogen peroxide (generated during reduction of oxygen in the potential range where cathode in a real fuel cell operates) are lowermost at composite material. In this respect, high activity of silver towards decomposition of the undesirable intermediate (HO$_2^-$) plays the key role. Another important factor is that the systems based on carbon nitrides retain their electrocatalytic activity even after continuous polarization in acidic media implying strong fixation of metal complexes to CN shells.

**Acknowledgments**

The support from the European Commission through the Graphene Flagship – Core 1 project [Grant number GA-696656] is highly appreciated.

**References**

[1] G. Merle, M. Wessling, K. Nijmeijer, J. Membr. Sci. 377, 1 (2011)

[2] M.A. Hickner, A.M. Herring, E.B. Coughlin, J. Polym. Sci., Part B: Polym. Phys. 51, 1727 (2013)

[3] N. Ramaswamy, S. Mukerjee, J. Phys. Chem. C 115, 18015 (2011)

[4] J.S. Spendelow, A. Wieckowski, Phys. Chem. Chem. Phys. 9, 2654 (2007) 2654

[5] M. Chatenet, L. Genies-Bultel, M. Aurousseau, R. Durand, F. Andolfatto, J. Appl. Electrochem. 32, 1131 (2002)

[6] L. Tammeveski, H. Erikson, A. Sarapuu, J. Kozlova, P. Ritslaid, V. Sammelselg, K. Tammeveski, Electrochem. Commun. 20, 15 (2012)

[7] F.W. Campbell, S.R. Belding, R. Baron, L. Xiao, R.G. Compton, J. Phys. Chem. C 113, 9053 (2009)

[8] A. Fazil, R. Chetty, Electroanalysis 26, 2380 (2014)

[9] J. Guo, A. Hsu, D. Chu, R. Chen, J. Phys. Chem. C. 114, 4324 (2010)

[10] J. Masa, A. Zhao, W. Xia, M. Muhler, W. Schuhmann, Electrochim. Acta 128, 271 (2014)

[11] K.-K. Türk, I. Kruusenberg, J. Mondal, P. Rauwel, J. Kozlova, L. Matisen, V. Sammelselg, K. Tammeveski, J. Electroanal. Chem. 756, 69 (2015)

[12] H.T. Chung, J.H. Won, P. Zelenay, Nat. Commun. 4:1922 (2013)

[13] V. Neburchilov, H. Wang, J.J. Martin, W. Qu, J. Power Sources 195, 1271 (2010)

[14] F. Bidault, D.J.L. Brett, P.H. Middleton, N.P. Brandon, J. Power Sources 187, 39 (2009)

[15] D. Geng, N. Ding, T.S.A. Hor, Z. Liu, X. Sun, Y. Zong, J. Mater. Chem. A 3, 1795 (2015)

[16] M. del Cueto, P. Ocón, J.M.L. Poyato, J. Phys. Chem. C 119, 2004 (2015)

[17] J. Liu, P. Song, Z. Ning, W. Xu, Electrocatalysis 6, 132 (2015)

[18] Z. Lin, G. Waller, Y. Liu, M. Liu, Ch.-P. Wong, Adv. Energy Mater. 2, 884 (2012)

[19] K. Sawai, N. Suzuki, Chem. Letters 33, 1540 (2004)

[20] K. Sawai, N. Suzuki, J. Electrochem. Soc. 151, A682 (2004)

[21] K. Sawai, Y.-S. Maeda, J. Electrochem. Soc. 155, B27 (2008)

[22] V. Di Noto, E. Negro, S. Polizzi, K. Vezzu, L. Toniolo, G. Cavinato, Int. J. Hydrogen Energy 39, 2812 (2014)

[23] E. Negro, S. Polizzi, K. Vezzu, L. Toniolo, G. Cavinato, V. Di Noto, Int. J. Hydrogen Energy 39, 2828 (2014)

[24] V. Di Noto, E. Negro, K. Vezzù, F. Bertasi, G. Nawn, Electrochem. Soc. Interface 24, 59 (2015)

[25] K Vezzù, A. Bach Delpeuch, E. Negro, S. Polizzi, G. Nawn, F. Bertasi, G. Pagot, K. Artyushkova, P. Atanassov, V. Di Noto, Electrochim. Acta 222, 1778 (2016)

[26] E. Negro, A. Bach Delpeuch, K. Vezzu', G. Nawn, F. Bertasi, A. Ansaldo, V. Pellegrini, B. Dembinska, S. Zoladek, K. Miecznikowski, I. Rutkowska, M. Skunik-Nuckowska, P.J. Kulesza, F. Bonaccorso, V. Di Noto, Chem. Mater. 30, 2651 (2018)

[27] Y. Liu, H. Wang, D. Lin, J. Zhao, C. Liu, J. Xie, Y. Cui, Nano Research 10, 1213 (2017)

[28] R. Liu, X. Yu, G. Zhang, S. Zhang, H. Cao, A. Dolbecq, P. Mialane, B. Keitad, L. Zhi, J. Mater. Chem. A 1, 11961 (2013)

[29] Y. Yang, Y. Zhou, J. Electroanal. Chem. 397, 271 (1995)

[30] B.B. Blizanac, P.N. Ross, N.M. Markovic, J. Phys. Chem. B 110, 4735 (2006)

[31] Y. Sun, M. Yang, J. Pan, P. Wang, W. Li, P. Wan, Electrochim. Acta 197, 68 (2016)

[32] Q. Wu, L. Jiang, L. Qi, L. Yuan, E. Wang, G. Sun, Electrochim. Acta 123, 167 (2014)

[33] F. Sun, G. Zhang, Y. Xu, Z. Chang, P. Wan, Y. Li, X. Sun, Electrochim. Acta 132, 136 (2014)

[34] Y. Wang, Y. Liu, X. Lu, Z. Li, H. Zhang, X. Cui, Y. Zhang, F. Shi, Y. Deng, Electrochem. Commun. 20, 171 (2012)

[35] Q. Yi, H. Chu, M. Tang, Z. Yang, Q. Chen, X. Liu, J. Electroanal. Chem. 739, 178 (2015)

[36] H.A. Miller, M. Bevilacqua, J. Filippi, A. Lavacchi, A. Marchionni, M. Marelli, S. Moneti, W. Oberhauser, E. Vesselli, M. Innocenti, F. Vizza, J. Mater. Chem. A 1, 13337 (2013)

[37] Y. Lu, N. Zhang, L. An, X. Li, D. Xia, J. Power Sources 240, 606 (2013)

[38] F.H.B. Lima, J.F.R. de Castro, Edson A. Ticianelli, J. Power Sources 161, 806 (2006)

[39] G.A. El-Nagar, I. Lauermann, R.M. Sarhan, C. Roth, Nanoscale, 10, 7304 (2018)

[40] E. Quesnel, F. Roux, F. Emieux, P. Faucherand, E. Kymakis, G. Volonakis, F. Giustino, B. Martín-García, I. Moreels, S. A. Gürsel, A. B. Yurtcan, V. Di Noto, A. Talyzin, I. Baburin, D. Tranca, G. Seifert, L. Crema, G. Speranza, V. Tozzini, P. Bondavalli, G. Pognon, C. Botas, D. Carriazo, G. Singh, T. Rojo, G. Kim, W. Yu, C. P. Grey, V. Pellegrini, 2D Mater 2, 030204

[41] D. Higgins, P. Zamani, A. Yu, Z. Chen, Energy Environ. Sci. 9, 357 (2016)

[42] Y. Li, Y. Li, E, Zhu, T. McLouth, C.-Y. Chiu, X. Huang, Yu. Huang, J. Am. Chem. Soc. 134, 12326 (2012)

[43] Y. Shao, S. Zhang, C. Wang, Z. Nie, J. Liu, Y. Wang, Y. Lin, J Power Sources 195, 4600 (2010)

[44] Y. Cui, S.I. Kundalwal, S. Kumar, Carbon 98, 313 (2016)

[45] D. He, K. Cheng, T. Peng, M. Pan, S. Mu, J. Mater. Chem. A 1, 2126 (2013)

[46] J. Jung, H.J. Park, J. Kim, S.H. Hur, J. Power Sources 248, 1156 (2014)

[47] J.-S. Lee, K. Jo, T. Lee, T. Yun, J. Cho, B.-Su. Kim, J. Mater. Chem. A 1, 9603 (2013)

[48] D. Ji, Y. Wang, S. Chen, Y. Zhang, L. Li, W. Ding, Z. Wei, J. Solid State Electrochem. (2018) https://doi.org/10.1007/s10008-018-3914-2

[49] A. Qaseem, F. Chen, X. Wu, N. Zhang, Z. Xia, Ag, J. Power Sources 370, 1 (2017)

[50] J.M. Linge, H. Erikson, A. Sarapuu, M. Merisalu, M. Rähn, L. Matisen, V. Sammelselg, K. Tammeveski, J. Electroanal. Chem. 794, 197 (2017)

[51] S. Hu, T. Han, C. Lin, W. Xiang, Y. Zhao, P. Gao, F. Du, X. Li, Y. Sun, Adv. Funct. Mater. 27, 1700041 (2017)

[52] K. Derwinska, K. Miecznikowski, R. Koncki, P.J. Kulesza, S. Glab, M.A. Malik, Electroanalysis 15, 1843 (2003)

[53] U. Schroder, F. Scholz, Inorg. Chem. 39, 1006 (2000)

[54] O. Antoine, R. Durand, J. Appl. Electrochem. 30, 839 (2000)

[55] T.J. Schmidt, U.A. Paulus, H.A. Gasteiger, R.J. Behm, J. Electroanal. Chem. 508, 41 (2001)

[56] M. Hepel, M. Tomkiewicz, J. Electrochem. Soc. 131, 1288 (1984)

[57] S. Zoladek, I.A. Rutkowska, M. Blicharska, K. Miecznikowski, W. Ozimek, J. Orlowska, E. Negro, V. Di Noto, P.J. Kulesza, Electrochim. Acta 233, 113 (2017)

**Figure captions**

**Fig. 1**. SEM and TEM images of the following materials: (**A,B**) GNP-800, (**C,D**) Ag/GNP800, (**E,F**) FeCN$_x$/GNP800, (**G,H**) AgFeCN$_x$/GNP800.

**Fig. 2.** Elemental EDX mapping of the following materials: (**A**) FeCN$_x$/GNP800, (**B**) AgFeCN$_x$/GNP800.

**Fig. 3.** XRD patterns of the following samples: (**a**) Ag/GNP800, (**b**) FeCN$_x$/GNP800 and (**c**) AgFeCN$_x$/GNP800.

**Fig. 4.** (**A**) Cyclic voltammetric responses recorded for electrode films (deposited on glassy carbon disk) of (**a**) Fe$_4$HCF$_3$/GNP, (**b**) Ag$_4$HCF/GNP and (**c**) GNP. Electrolyte: nitrogen-saturated 0.5 mol dm$^{-3}$ K$_2$SO$_4$. Scan rate: 50 mV s$^{-1}$. (**B**) Voltammograms recorded for catalytic films of (**a**) Ag/GNP800-C, (**b**) FeCN$_x$/GNP800-C, (**c**) AgFeCN$_x$/GNP800-C and (**d**) GNP800-C. Electrolyte: nitrogen-saturated 0.1 mol dm$^{-3}$ KOH. Scan rate: 10 mV s$^{-1}$.

**Fig. 5.** Normalized (background subtracted) rotating disk (**A**) and ring (**B**) voltammograms recorded during oxygen reduction at the films of (**a**) Ag/GNP800-C, (**b**) FeCN$_x$/GNP800-C, (**c**) AgFeCN$_x$/GNP800-C (**d**) GNP800-C and (**e**) Pt/C (E-tek) in the oxygen-saturated 0.1 mol dm$^{-3}$ KOH at the scan rate of 10 mV s$^{-1}$ and rotation rate of 1600 rpm. Formation of the hydrogen peroxide intermediate (**C**) and number of exchanged electrons (**D**) during oxygen reduction under conditions of RRDE voltammetric experiments.

**Fig. 6.** Cyclic voltammograms recorded for catalytic films of (**A**) Ag/GNP800-C, (**B**) FeCN$_x$/GNP800-C and (**C**) AgFeCN$_x$/GNP800-C recorded in 0.1 mol dm$^{-3}$ KOH (at 100 mV s$^{-1}$) before (dashed lines) and after (solid lines) preconditioning (*P*) in 0.1 mol dm$^{-3}$ HClO$_4$.

**Fig. 7.** Normalized (background subtracted) rotating disk and ring voltammograms recorded during oxygen reduction at the films of (**A,A'**) Ag/GNP800-C (**B, B'**) FeCN$_x$/GNP800-C and (**C, C'**) AgFeCN$_x$/GNP800-C in the oxygen-saturated 0.1 mol dm$^{-3}$ KOH at the scan rate of 10 mV s$^{-1}$ and rotation rate of 1600 rpm before (dashed lines) and after preconditioning in 0.1 mol dm$^{-3}$ HClO$_4$ (abbreviated as ***P***, solid lines).

**Fig. 8.** (**A**) TEM image of AgFeCN$_x$/GO800. Normalized rotating disk (**B**) and ring (**C**) voltammograms recorded during oxygen reduction at the films of (**a**) AgFeCN$_x$/GNP800-C and (**b**) AgFeCN$_x$/GO800-C in the oxygen-saturated 0.1 mol dm$^{-3}$ KOH at the scan rate of 10 mV s$^{-1}$ and rotation rate of 1600 rpm.

**Figures**

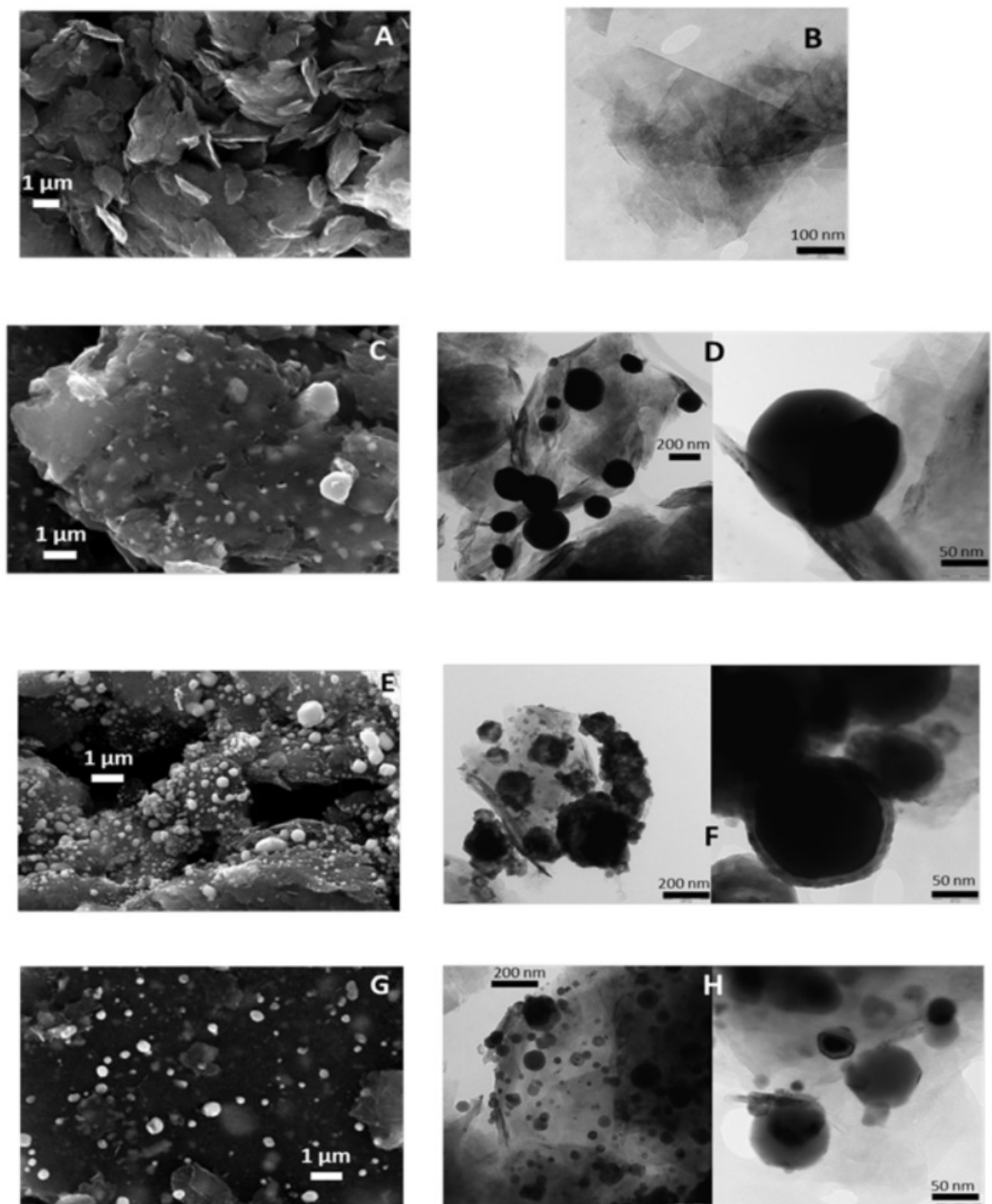

**Fig. 1**. SEM and TEM images of the following materials: (**A,B**) GNP-800, (**C,D**) Ag/GNP800, (**E,F**) FeCN$_x$/GNP800, (**G,H**) AgFeCN$_x$/GNP800.

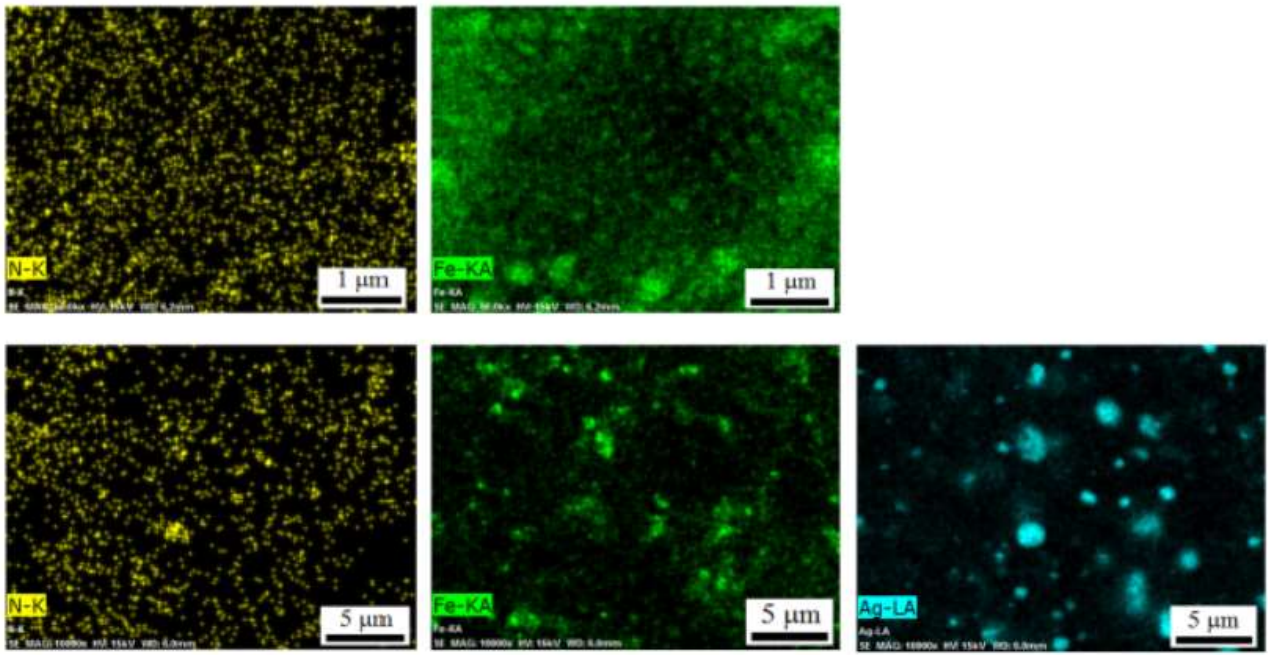

**Fig. 2.** Elemental EDX mapping of the following materials: **(A)** FeCN$_x$/GNP800, **(B)** AgFeCN$_x$/GNP800.

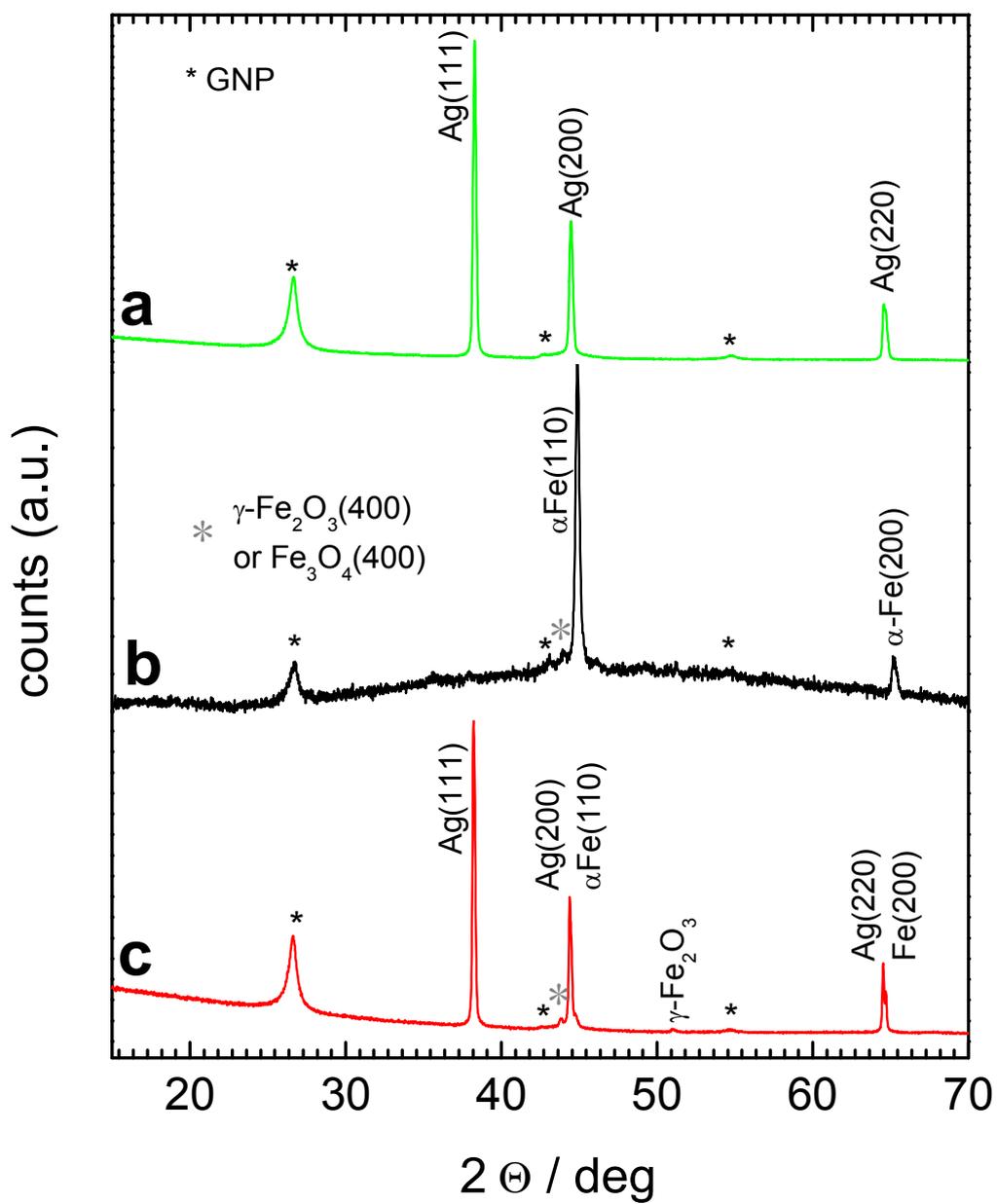

**Fig. 3.** XRD patterns of the following samples: (**a**) Ag/GNP800, (**b**) FeCN$_x$/GNP800 and (**c**) AgFeCN$_x$/GNP800.

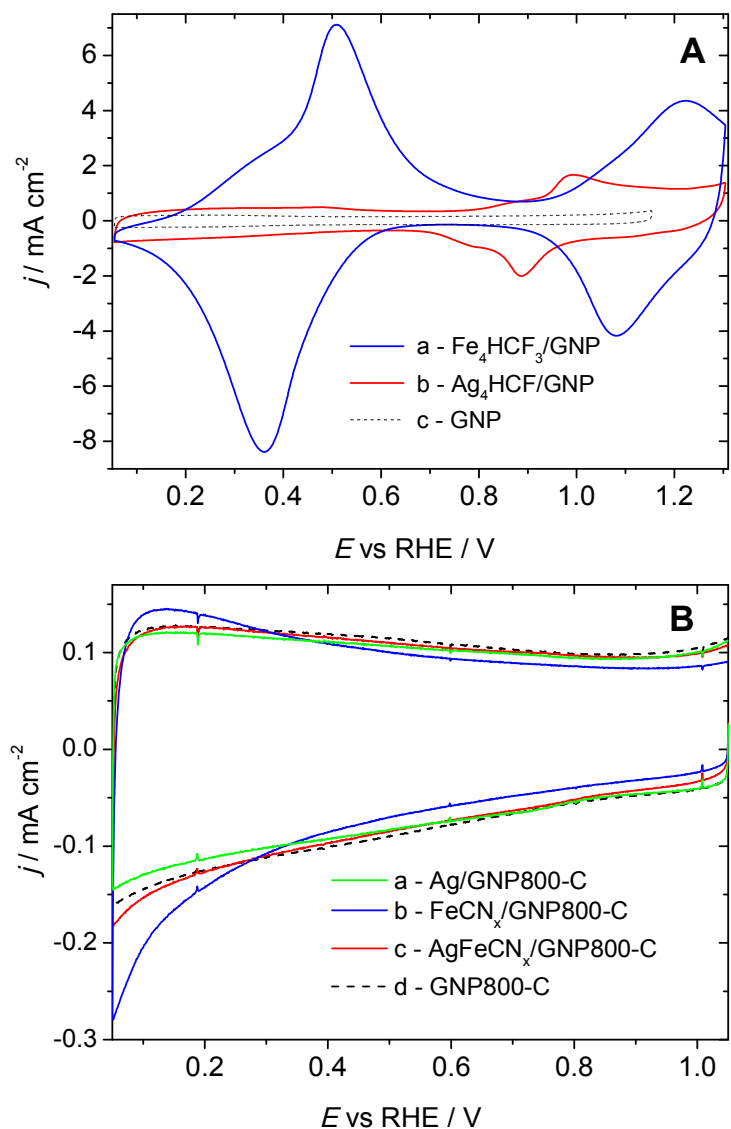

**Fig. 4.** (**A**) Cyclic voltammetric responses recorded for electrode films (deposited on glassy carbon disk) of (**a**) $Fe_4HCF_3$/GNP, (**b**) $Ag_4HCF$/GNP and (**c**) GNP. Electrolyte: nitrogen-saturated 0.5 mol dm$^{-3}$ $K_2SO_4$. Scan rate: 50 mV s$^{-1}$. (**B**) Voltammograms recorded for catalytic films of (**a**) Ag/GNP800-C, (**b**) $FeCN_x$/GNP800-C, (**c**) $AgFeCN_x$/GNP800-C and (**d**) GNP800-C. Electrolyte: nitrogen-saturated 0.1 mol dm$^{-3}$ KOH. Scan rate: 10 mV s$^{-1}$.

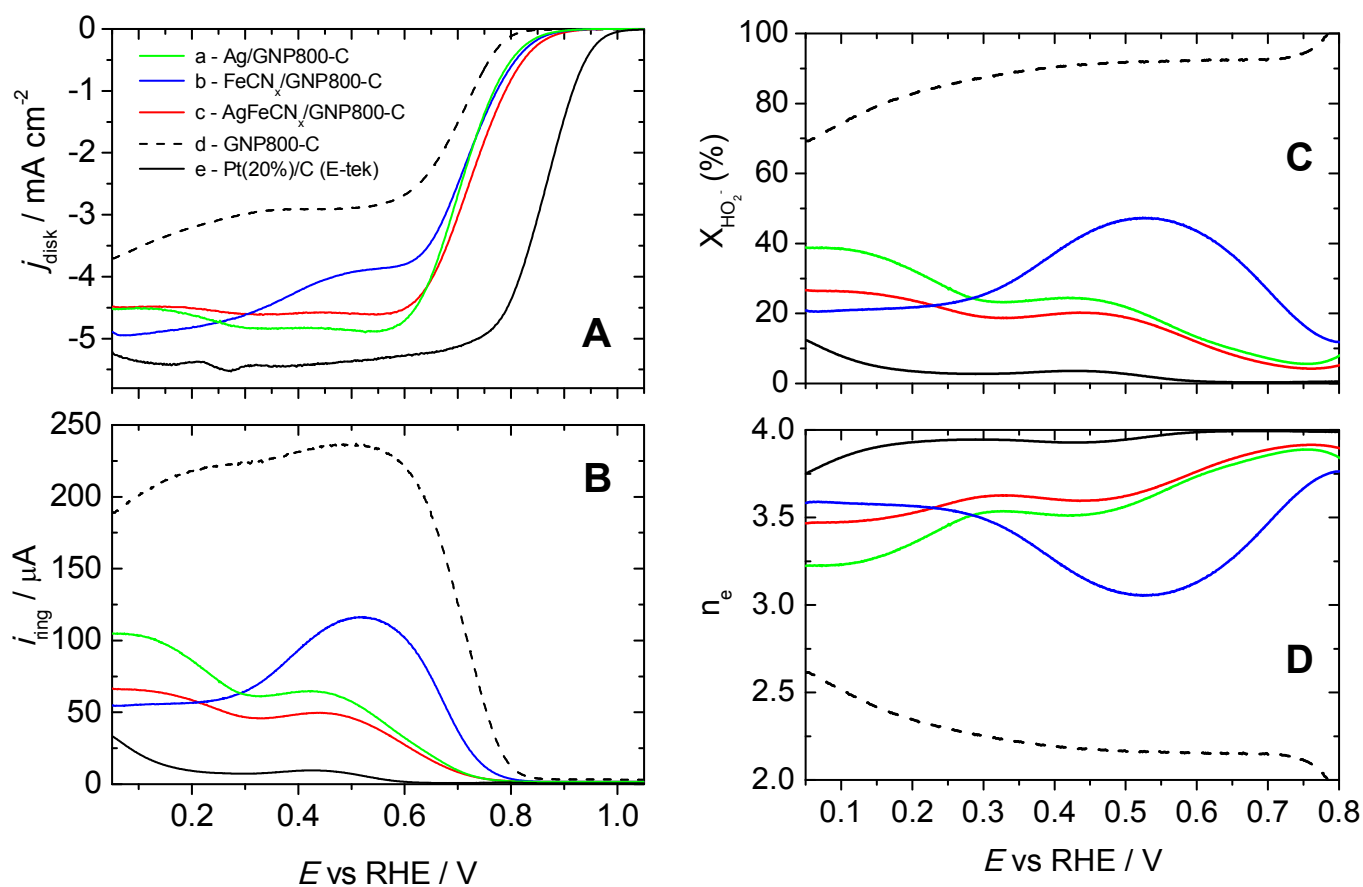

**Fig. 5.** Normalized (background subtracted) rotating disk (**A**) and ring (**B**) voltammograms recorded during oxygen reduction at the films of (**a**) Ag/GNP800-C, (**b**) FeCN$_x$/GNP800-C, (**c**) AgFeCN$_x$/GNP800-C (**d**) GNP800-C and (**e**) Pt/C (E-tek) in the oxygen-saturated 0.1 mol dm$^{-3}$ KOH at the scan rate of 10 mV s$^{-1}$ and rotation rate of 1600 rpm. Formation of the hydrogen peroxide intermediate (**C**) and number of exchanged electrons (**D**) during oxygen reduction under conditions of RRDE voltammetric experiments.

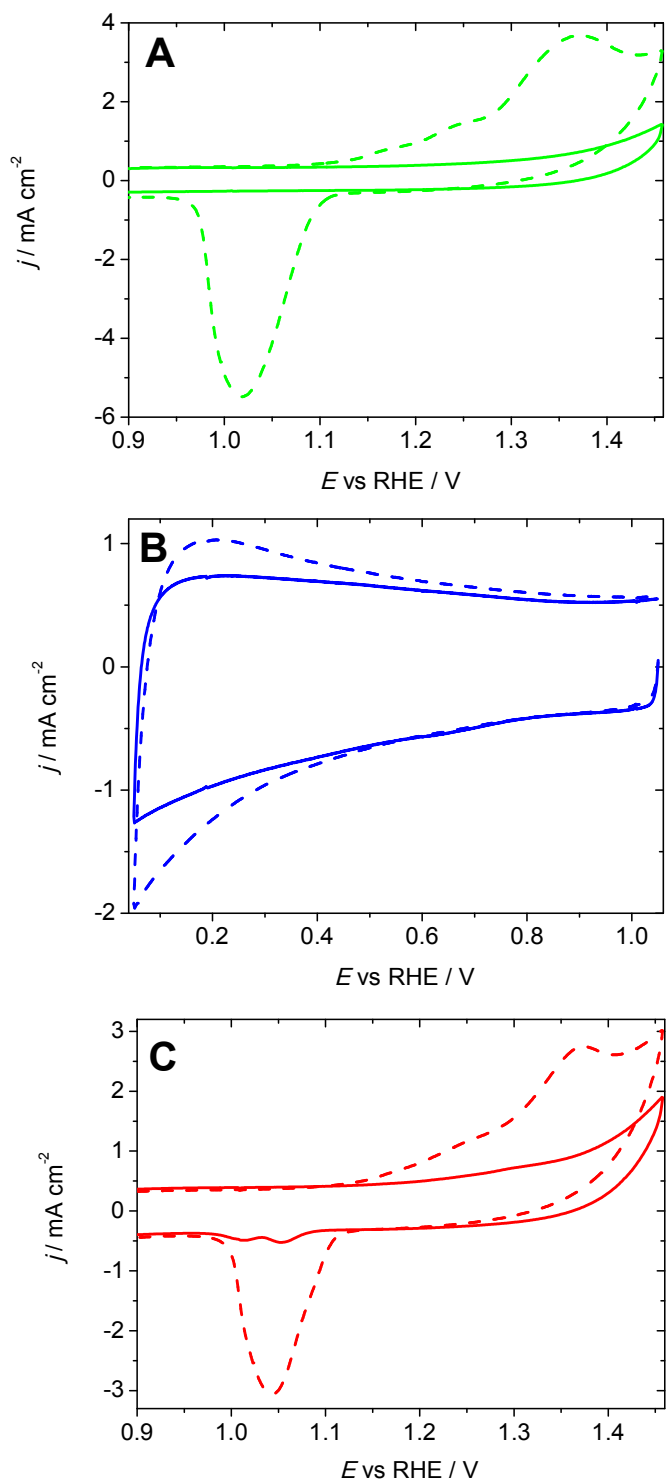

**Fig. 6.** Cyclic voltammograms recorded for catalytic films of (**A**) Ag/GNP800-C, (**B**) FeCN$_x$/GNP800-C and (**C**) AgFeCN$_x$/GNP800-C recorded in 0.1 mol dm$^{-3}$ KOH (at 100 mV s$^{-1}$) before (dashed lines) and after (solid lines) preconditioning (**P**) in 0.1 mol dm$^{-3}$ HClO$_4$.

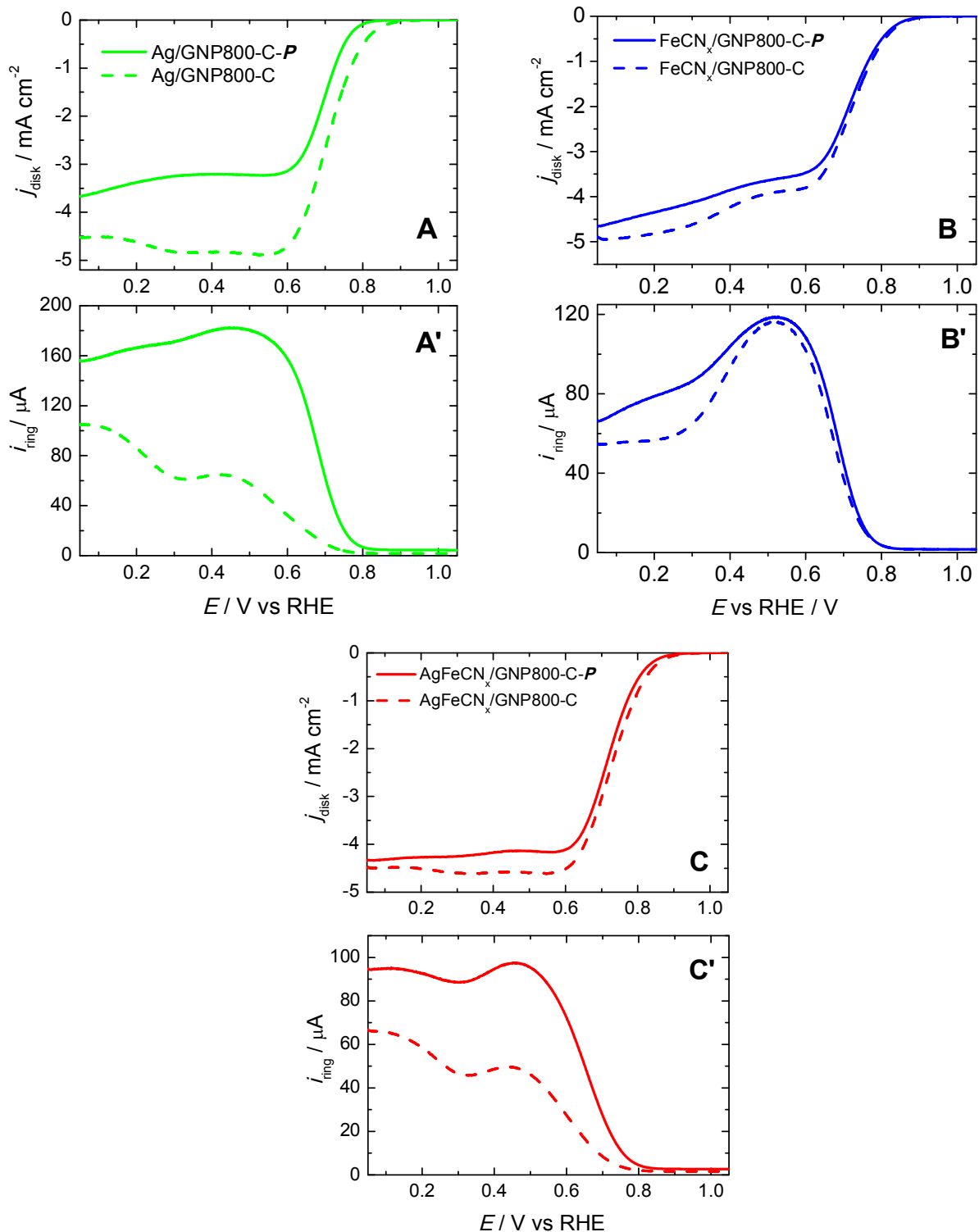

**Fig. 7.** Normalized (background subtracted) rotating disk and ring voltammograms recorded during oxygen reduction at the films of (**A,A'**) Ag/GNP800-C (**B, B'**) FeCN$_x$/GNP800-C and (**C, C'**) AgFeCN$_x$/GNP800-C in the oxygen-saturated 0.1 mol dm$^{-3}$ KOH at the scan rate of 10 mV s$^{-1}$ and rotation rate of 1600 rpm before (dashed lines) and after preconditioning in 0.1 mol dm$^{-3}$ HClO$_4$ (abbreviated as ***P***, solid lines).

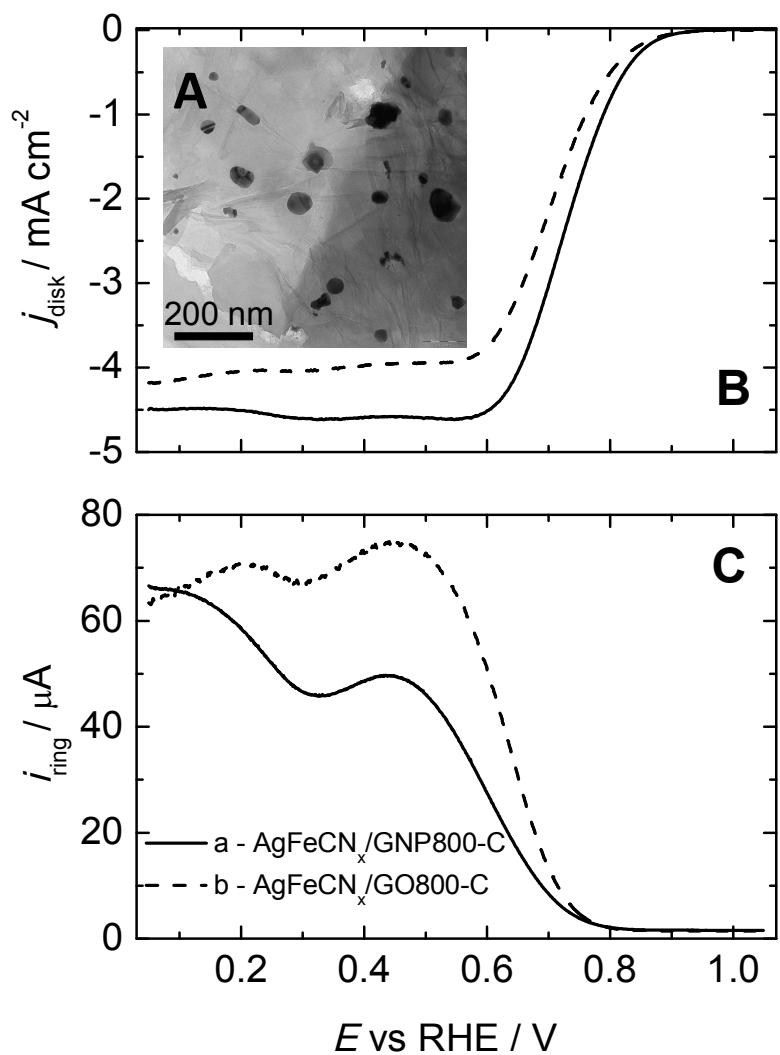

**Fig. 8.** (**A**) TEM image of AgFeCN$_x$/GO800. Normalized rotating disk (**B**) and ring (**C**) voltammograms recorded during oxygen reduction at the films of (**a**) AgFeCN$_x$/GNP800-C and (**b**) AgFeCN$_x$/GO800-C in the oxygen-saturated 0.1 mol dm$^{-3}$ KOH at the scan rate of 10 mV s$^{-1}$ and rotation rate of 1600 rpm.